\documentclass[prb,longbibliography,showpacs,twocolumn,superscriptaddress,amsmath,amssymb,verbatim]{revtex4-1}

\usepackage{graphicx}
\usepackage{lmodern}
\usepackage{epstopdf}
\usepackage{dcolumn}
\usepackage{bm}
\usepackage{subfigure}
\usepackage{amsmath} 

\usepackage[pdftex,colorlinks=true,citecolor=blue,linkcolor=blue,urlcolor=blue,bookmarks=true]{hyperref}
\usepackage[sort&compress]{natbib}

\usepackage{color}
\usepackage{textcomp}
\definecolor{red}{rgb}{1,0,0}

\definecolor{blue}{rgb}{0,0,1}

\definecolor{green}{rgb}{0,1,0}

\begin{document}
\preprint{APS}

\title{Magnetic ground state of the frustrated spin-1/2 chain compound $\beta$-TeVO$_4$ at high magnetic fields}

\author{M. Pregelj}
\email{matej.pregelj@ijs.si}
\affiliation{Jo\v{z}ef Stefan Institute, Jamova 39, 1000 Ljubljana, Slovenia}
\author{A. Zorko}
\affiliation{Jo\v{z}ef Stefan Institute, Jamova 39, 1000 Ljubljana, Slovenia}
\author{M. Klanj\v{s}ek}
\affiliation{Jo\v{z}ef Stefan Institute, Jamova 39, 1000 Ljubljana, Slovenia}
\author{O. Zaharko}
\affiliation{Laboratory for Neutron Scattering and Imaging, PSI, CH-5232 Villigen, Switzerland}
\author{J. S. White}
\affiliation{Laboratory for Neutron Scattering and Imaging, PSI, CH-5232 Villigen, Switzerland}
\author{O. Prokhnenko}
\affiliation{Helmholtz-Zentrum Berlin für Materialien und Energie, Hahn-Meitner-Platz 1, 14109 Berlin, Germany}
\author{M. Bartkowiak}
\affiliation{Helmholtz-Zentrum Berlin für Materialien und Energie, Hahn-Meitner-Platz 1, 14109 Berlin, Germany}
\author{H. Nojiri}
\affiliation{Institute for Materials Research, Tohoku University, Sendai 980-8577, Japan}
\author{H. Berger}
\affiliation{Ecole polytechnique f\'{e}d\'{e}rale de Lausanne, CH-1015 Lausanne, Switzerland}
\author{D. Ar\v{c}on}
\affiliation{Jo\v{z}ef Stefan Institute, Jamova 39, 1000 Ljubljana, Slovenia}
\affiliation{Faculty of Mathematics and Physics, University of Ljubljana, Jadranska c. 19, 1000 Ljubljana, Slovenia}

\date{\today}

\begin{abstract}

Frustrated spin-1/2 chains, despite the apparent simplicity, exhibit remarkably rich phase diagram comprising vector-chiral (VC), spin-density-wave (SDW) and multipolar/spin-nematic phases as a function of the magnetic field.
Here we report a study of $\beta$-TeVO$_4$, an archetype of such compounds, based on magnetization and neutron diffraction measurements up to 25\,T. 
We find the transition from the helical VC ground state to the SDW state at $\sim$3\,T for the magnetic field along the $a$ and $c$ crystal axes, and at $\sim$9\,T for the field along the $b$ axis.
The high-field (HF) state, existing above $\sim$18\,T, i.e., above $\sim$1/2 of the saturated magnetization, is an incommensurate magnetically ordered state and not the spin-nematic state, as theoretically predicted for the isotropic frustrated spin-1/2 chain.
The HF state is likely driven by sizable interchain interactions and symmetric intrachain anisotropies uncovered in previous studies. 
Consequently, the potential existence of the spin nematic phase in $\beta$-TeVO$_4$ is limited to a narrow field range, i.e., a few tenths of a tesla bellow the saturation of the magnetization, as also found in other frustrated spin-1/2 chain compounds.

\end{abstract}

\maketitle

\section{Introduction}

Frustrated spin-1/2 chains have been lately drawing considerable attention due to an intriguing magnetic quadrupolar,\cite{andreev1984spin,blume1969biquadratic,lacroix2011introduction} i.e., spin-nematic, phase that is predicted to occur in the applied magnetic field just before the magnetization gets saturated.\cite{hikihara2008vector,sudan2009emergent}
This remarkable state features no regular, i.e., dipolar magnetic order, but rather consists of ordered magnetic quadrupoles that are formed out of bound magnon pairs, which condense at the bonds between the neighboring spins.\cite{chubukov1991chiral, shannon2006nematic,zhitomirsky2010magnon}
As such, this novel state of matter is notoriously difficult to detect experimentally. 
The excitations of the spin-nematic order, however, break the bound magnon pairs and thus couple to the magnetic field.\cite{sato2009nmr,starykh2014excitations}
The spin-nematic phase is, therefore, expected to be reflected either in a missing magnetization fraction\cite{orlova2017nuclear} or in a particular kind of magnetic-excitation dispersion.\cite{sato2009nmr,sato2011field,starykh2014excitations,onishi2015magnetic,furuya2017angular}
Unfortunately, the most studied frustrated spin-1/2 chain candidate LiCuVO$_4$ has a saturation field of $\sim$40\,T,\cite{svistov2011new} which severely hinders the applicable experimental techniques.
The experimental reports of the spin-nematic phase are thus limited\cite{svistov2011new,mourigal2012evidence,buttgen2014search,orlova2017nuclear} and so far have not provided an indisputable proof of its existence in physical systems.

\begin{figure}[!]
\centering
\includegraphics[width=\columnwidth]{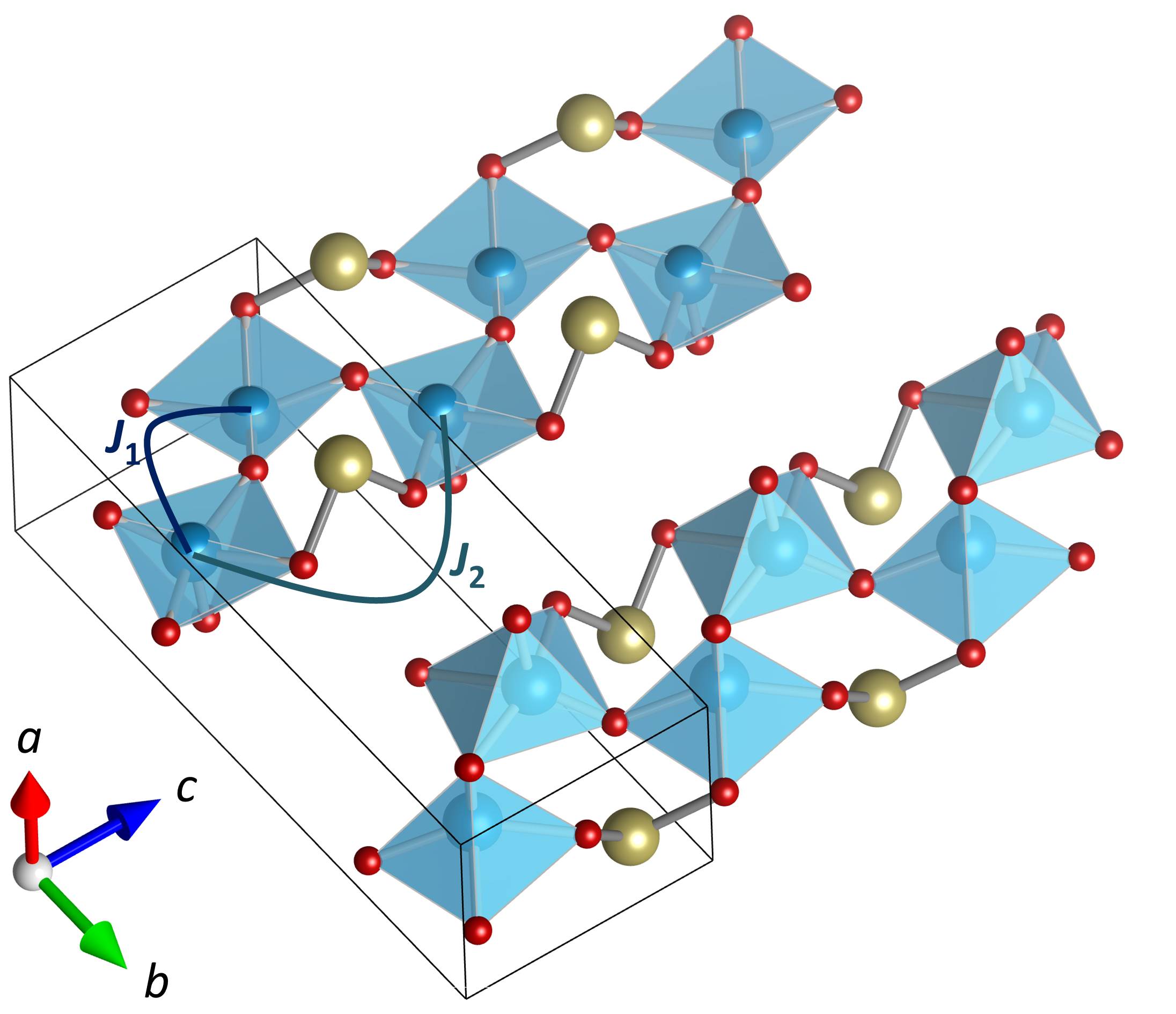}
\caption{The crystal structure of $\beta$-TeVO$_4$. Large blue spheres are vanadium atoms, small red spheres are oxygen atoms, and light brown spheres are tellurium atoms, whereas $J_1$ and $J_2$ denote the nearest- and next-nearest-neighbor exchange pathways. }
\label{fig-struc}
\end{figure}

\begin{figure}[!]
\centering
\includegraphics[width=0.9\columnwidth]{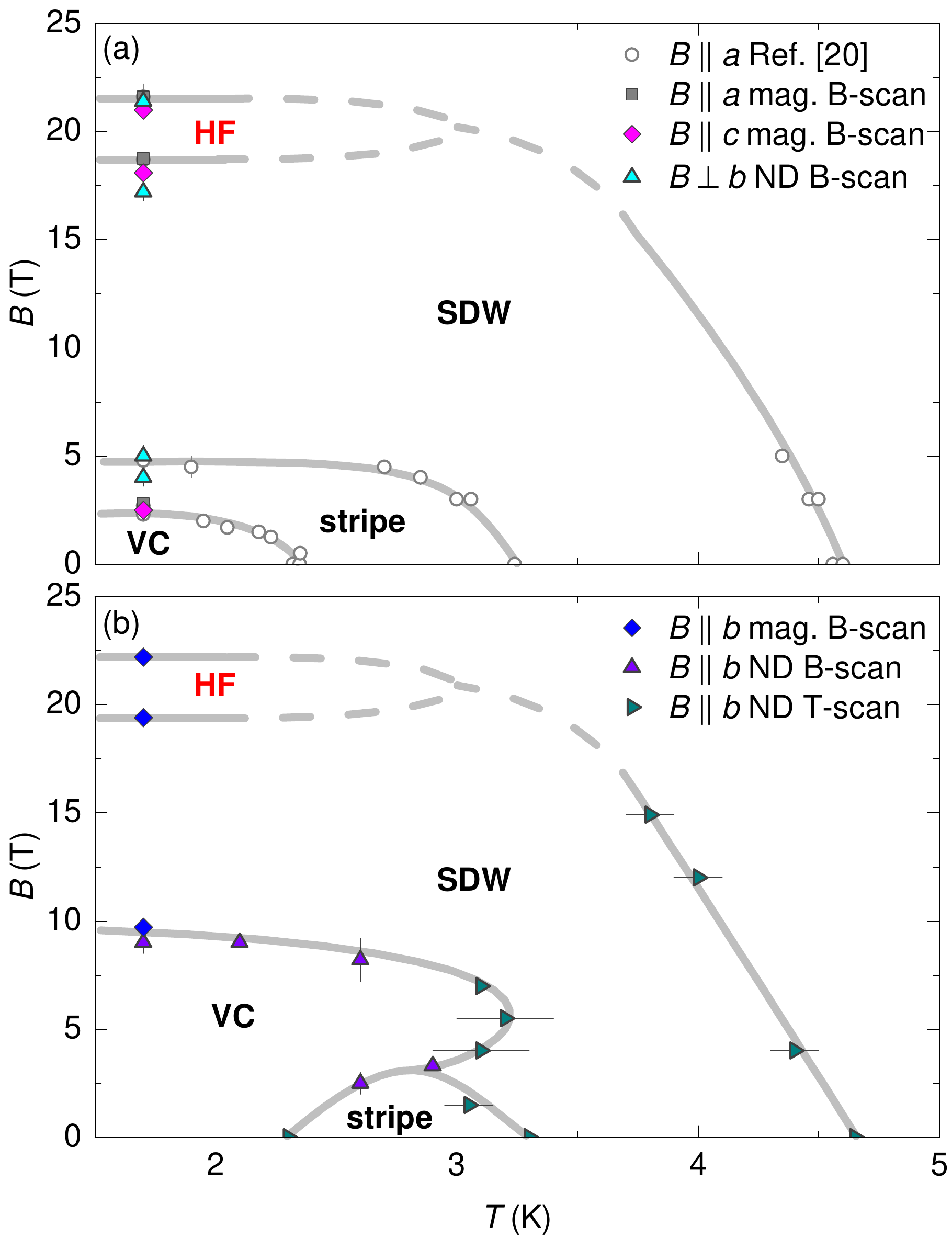}
\caption{Magnetic phase diagrams derived for the magnetic field applied (a) perpendicular and (b) parallel to the $b$ axis.}
\label{fig-PD}
\end{figure}

Here we focus on $\beta$-TeVO$_4$,\cite{meunier1973oxyde,savina2011magnetic} which has been recently recognized as a very good realization of a spin-1/2 frustrated ferromagnetic chain.\cite{pregelj2015spin,savina2015study,weickert2016magnetic,pregelj2018coexisting}
In this compound, the V$^{4+}$ ($S$\,=\,1/2) magnetic ions are coupled by ferromagnetic nearest-neighbor $J_1$\,$\approx$\,$-$38\,K and antiferromagnetic next-nearest-neighbor $J_2$\,$\approx$\,$-J_1$ exchange interactions mediated by oxygen and oxygen-tellurium-oxygen superexchange pathways, respectively (Fig.\,\ref{fig-struc}).
The compound undergoes a long-range magnetic ordering below $T_{N1}$\,=\,4.6\,K and the experimentally derived phase diagram [Fig.\,\ref{fig-PD}(a)] almost exactly matches the theoretically predicted diagram.\cite{pregelj2015spin}
Namely, the helical ground state, denoted also vector-chiral (VC) state, is at $\sim$3\,T succeeded by the spin-density-wave (SDW) state that exhibits additional stripe modulation up to $\sim$5\,T and is at $\sim$18.7\,T followed by another phase, whose properties are not yet determined.
According to the theory, this phase could be a spin-nematic phase, persisting up to saturation of the magnetization at $\sim$21.5\,T.
The fact that the high-field (HF) phase develops already at moderately high fields of $\sim$18.7\,T opens the possibility to explore this phase with neutron scattering experiments that could not have been done for other candidate compounds. 
These experiments should finally provide firm experimental evidence either in favor or against the existence of the spin-nematic order in $\beta$-TeVO$_4$.

In this work we combine neutron diffraction experiments on $\beta$-TeVO$_4$ in applied magnetic fields up to 25\,T with an in-depth high-field magnetization study.
The data reveal that the HF phase exhibits long-range incommensurate magnetic order and is thus not the spin-nematic phase that is theoretically predicted for the isotropic frustrated spin-1/2 chain.
We argue that this is most likely due to sizable interchain exchange interactions and strong symmetric anisotropy of the intrachain interactions, which induce the stripe modulation in the narrow region of the SDW phase\cite{pregelj2015spin} and probably also help stabilizing the incommensurate magnetic order at high magnetic fields.

\section{Experimental}

The single-crystal samples were grown from TeO$_2$ and VO$_2$ powders by chemical vapor transport reaction, using a two-zone furnace and TeCl$_4$ as a transport agent.\cite{pregelj2015spin,weickert2016magnetic,pregelj2018coexisting}
Magnetization measurements in pulsed magnetic fields up to 25\,T were performed on a 1.6$\times$0.7$\times$0.8\,mm$^3$ single-crystal at the High Magnetic Field Laboratory, Institute for Materials Research, Sendai, Japan.
This sample was cut from a larger sample oriented by neutron diffraction.
Neutron diffraction measurements were performed on a 2$\times$3$\times$4\,mm$^3$ single-crystal.
Measurements in magnetic field up to 15\,T applied along the $b$ axis were performed on the triple-axis-spectrometer TASP at the Paul Scherrer Institute (PSI), Villigen, Switzerland.
An analyzer was used to reduce the background, while the standard ILL orange cryostat was used for cooling.
Measurements in magnetic field applied up to 25\,T in the $ac$ plane were performed on the HFM/EXED instrument at the Helmholtz Zentrum Berlin, Berlin, Germany, which is the only existing instrument/facility that allows for neutron scattering experiments in static magnetic fields up to 26\,T.\cite{prokhnenko2015time, prokhnenko2017hfm}
Due to the complex construction of the HFM/EXED hybrid magnet, the magnetic field can be applied only horizontally with the magnet opening of $\sim$30$^\circ$.
Hence, when probing the (0.2~0~$-$0.42) magnetic reflection the magnetic field cannot be applied exactly along the $a$ axis, but rather rotated towards the $c$ axis by $\sim$30$^\circ$.
To cover the broad region of the reciprocal space ($h$, 0, $l$), where $0\leq h \leq1$ and $-1\leq l \leq0$, the sample was rotated around the $b$ axis from the $a$ towards the $c$ axis in steps of $\sim$10$^\circ$.
The maps of the reciprocal space shown in Fig.\,\ref{fig-NDmaps} are thus obtained by combining data sets measured at different field orientations within the $ac$ plane.
We note that the HFM/EXED instrument does not have beam collimators and employs a conical cryostat, which puts a substantial amount of the material in the neutron beam, leading to a significant background that is also angle-dependent.

\section{Results}

\subsection{Magnetization measurements}

\begin{figure}[!]
\centering
\includegraphics[width=\columnwidth]{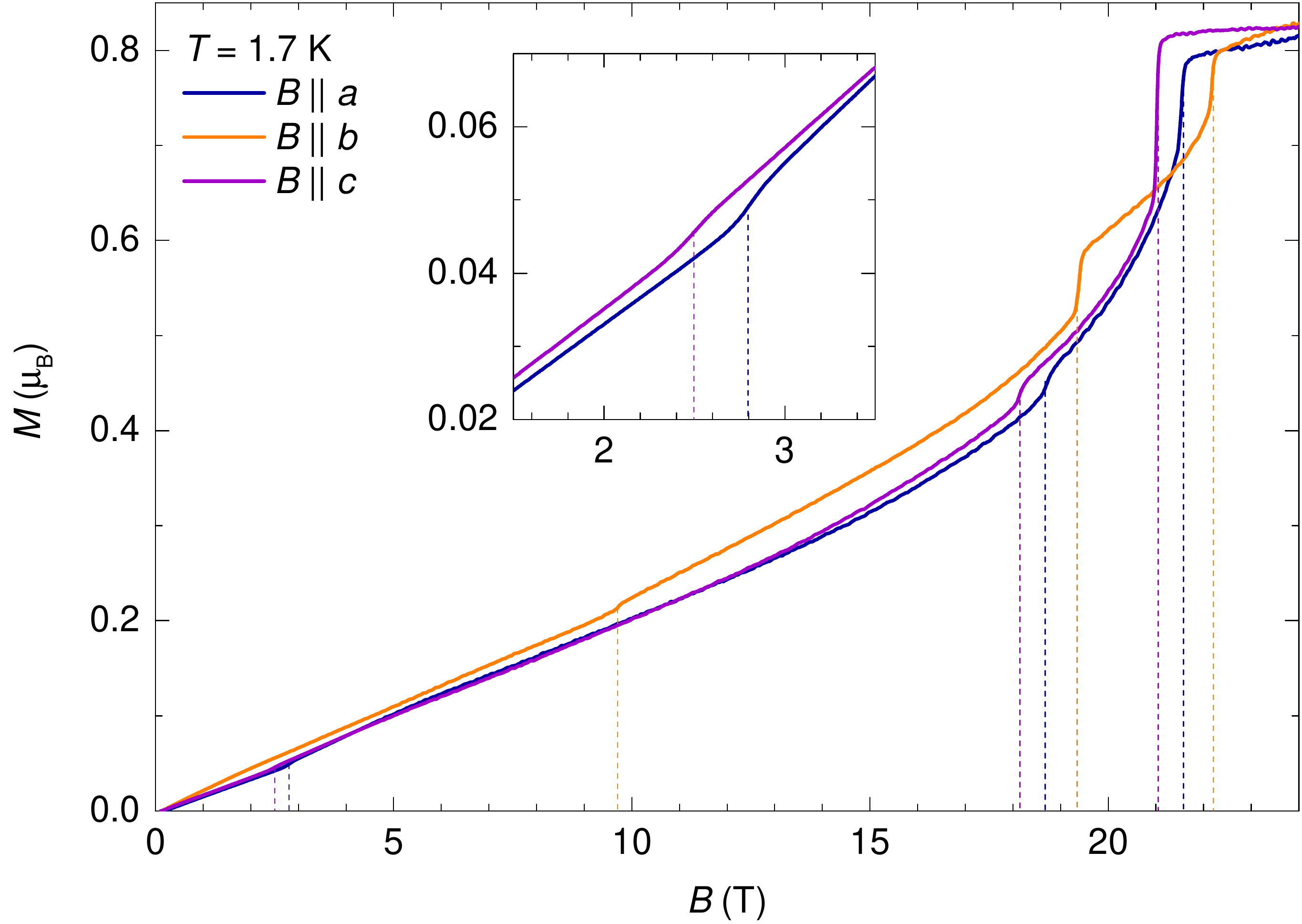}
\caption{Magnetization measured for the magnetic field applied along the $a$, $b$, and $c$ crystallographic axes. Inset: enlarged plot for easier visualization of the low-field transition for $B||a$ and $B||c$.}
\label{fig-mag}
\end{figure}

Magnetization measurements were performed in pulsed magnetic field at 1.7\,K to identify low-temperature field-induced magnetic transitions.
The data obtained for magnetic field ($B$) applied along all three crystallographic axes ($a$, $b$, and $c$) are shown in Fig.\,\ref{fig-mag} and summarized in Fig.\,\ref{fig-PD}.
The magnetization curves exhibit three anomalies, indicating magnetic transitions from the VC ground state to the SDW state, from the SDW to the HF state, and finally into the fully saturated state.
The response for $B||a$ and $B||c$ is rather similar, i.e., the anomalies for $B||a$ ($B||c$) are found at 2.8 (2.5), 18.7 (18.1) and 21.5 (21.0)\,T.
On the other hand, for $B||b$ the low-field anomaly is shifted to significantly higher field, i.e., it occurs at 9.7\,T, whereas the other two are found at only slightly higher fields than for the other two directions, i.e., at 19.4 and 22.2\,T.
This is not surprising when considering the strong magnetic anisotropy highlighted in preceding studies.\cite{pregelj2015spin,pregelj2016exchange,weickert2016magnetic,pregelj2018coexisting,pregelj2019elementary}

\subsection{Neutron diffraction}

In order to explore the magnetic phase diagram in more detail, we performed neutron diffraction measurements in the presence of the applied magnetic field.

\subsubsection{Magnetic field along the $b$ axis}

\begin{figure*}[!]
\centering
\includegraphics[width=17cm]{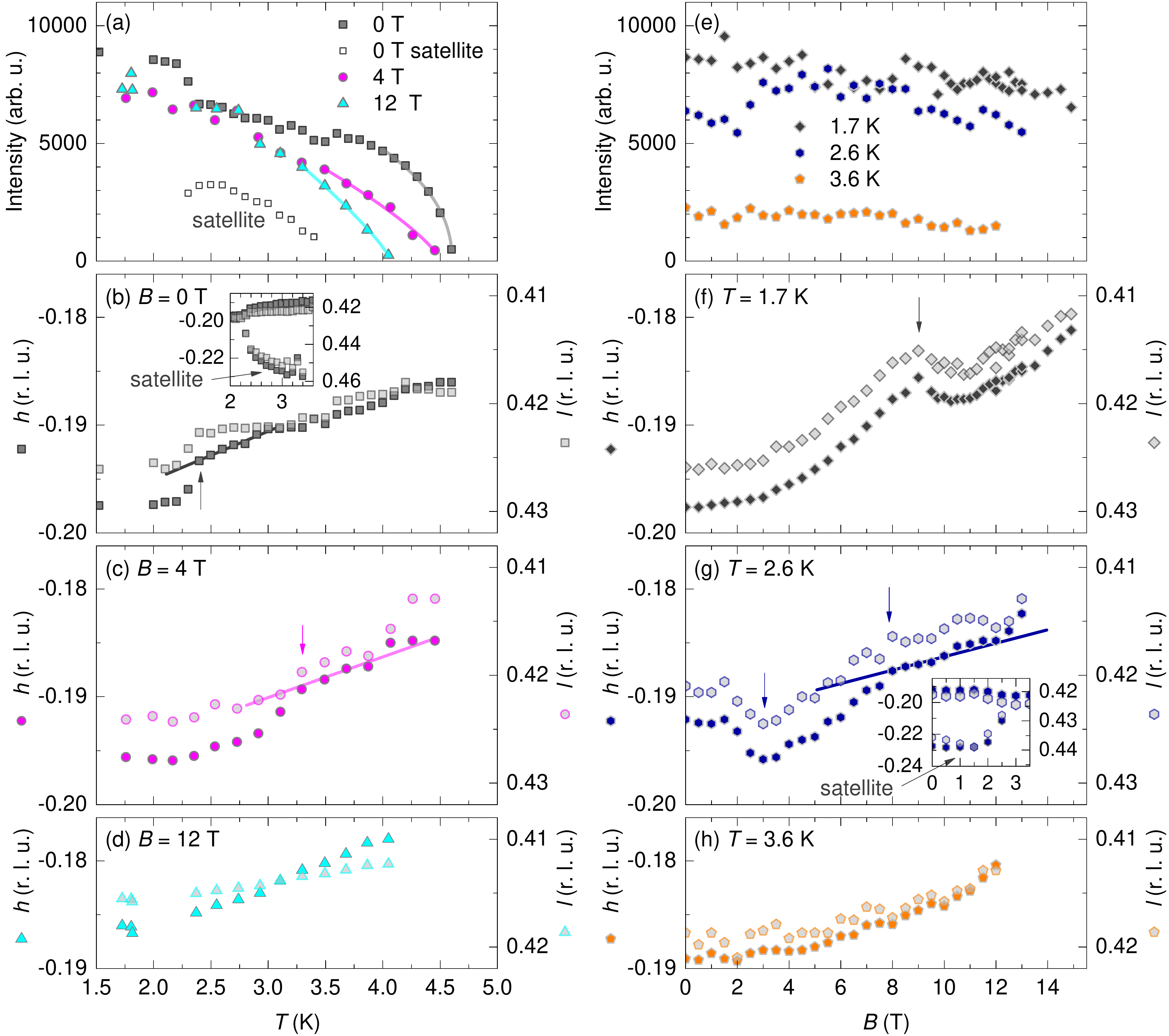}
\caption{Neutron diffraction measurements of the ($-$0.2~0~0.42) magnetic reflection and its satellite (existing only in the spin-stripe modulated phase) for magnetic field applied along the $b$ axis. The position of the magnetic reflection directly determines the magnetic wave vector. (a) The temperature dependencies of the intensity of the ($-$0.2~0~0.42) reflection and its satellite in different magnetic fields. Solid lines are fits of the critical behavior $\propto$\,$(T_{N1}-T)^{2\beta}$, where $\beta$ is the critical exponent. (b)-(d) The temperature dependence of the measured $h$ and $l$ components ($k$\,=\,0) of the magnetic wave vector at 0, 4 and 12\,T. Inset in (b) shows also the wave vector corresponding to the satellite reflection in the spin-stripe modulation.  (e) The magnetic field dependencies of the ($-$0.2~0~0.42) reflection intensity at different temperatures. (f)-(h) The magnetic field dependence of the $h$ and $l$ components ($k$\,=\,0) of the magnetic wave vector measured at 1.7, 2.6, and 3.6\,K. Inset in (g) shows also the wave vector corresponding to the satellite reflection in the spin-stripe modulation. Solid lines are guides to the eye, allowing for easier identification of the phase transitions indicated by the vertical arrows.}
\label{fig-NDb}
\end{figure*}

First, we measured temperature dependencies of the second strongest magnetic reflection, i.e., the ($-$0.2~0~0.42) reflection that also coincides with the magnetic wave vector, at several fixed magnetic fields applied along the $b$ axis [Fig.\,\ref{fig-NDb}(a)-(d)].
The intensity data reveal a gradual reduction of the magnetic-ordering transition temperature $T_{N1}$, which reduces from 4.60(2)\,K at 0\,T to 4.10(2)\,K at 12\,T [Fig.\,\ref{fig-NDb}(a)].
Considering that in the vicinity of the magnetic transition the intensity of the magnetic reflection is proportional to the square of the magnetic order parameter, i.e., $\propto$\,$(T_{N1}$$-$$T)^{2\beta}$, we find an increase of the critical exponent $\beta$, from 0.22(2) at 0\,T to 0.41(2) at 12\,T.
This suggests a crossover from two-dimensional XY (2D planar, $\beta$\,$\approx$\,0.23) towards three-dimensional Heisenberg (3D isotropic, $\beta$\,$\approx$\,0.36) universality class,\cite{taroni2008universal,pelissetto2002critical,chaikin2000principles} implying that with increasing field the intrachain correlations are enhanced and become more isotropic.
In the absence of the magnetic field, the occurrence of the satellite magnetic reflection at $T_{N2}$\,=\,3.35(5)\,K indicates the establishment of the spin-stripe modulation, while the anomaly at $T_{N3}$\,=\,2.35(2)\,K, where the satellite reflection merges with the main one, indicates the establishment of the VC phase.
The two transitions are also associated with the changes of the magnetic wave vector.
The latter first shifts linearly from $T_{N1}$ to $T_{N3}$ and then suddenly jumps and locks to the low temperature value [Fig.\,\ref{fig-NDb}(b)]. 
In contrast, the temperature dependence measured at 4\,T does not exhibit any significant anomaly in the intensity of the magnetic reflection as well as no satellite reflection is observed, indicating that spin-stripe modulation does not develop at this field.
Nevertheless, the position of the magnetic reflection exhibits a weak anomaly at 3.50(5)\,K, which is followed by gradual flattening of its temperature dependence [Fig.\,\ref{fig-NDb}(c)], eventually locking to the low-temperature value.
This implies that also for this field the system undergoes the transition into the VC phase.
Finally, the temperature dependence measured at 12\,T [Fig.\,\ref{fig-NDb}(d)] shows no anomaly in the magnetic peak intensity nor in its position, indicating that for this field the system persists in the SDW state down to 1.7\,K.

In the next step, we performed complementary field-dependence measurements at fixed temperatures [Fig.\,\ref{fig-NDb}(e)-(h)].
At 1.7 and 3.6\,K the intensity of the ($-$0.2~0~0.42) reflection exhibits no anomalies up to 15\,T, implying that the projection of the magnetic order to the plane perpendicular to the magnetic wave vector does not change significantly at these temperatures [Fig.\,\ref{fig-NDb}(e)].
The peak position measured at 1.7\,K, however, exhibits a clear local maximum at 9.0(2)\,T that must be associated with the VC-to-SDW magnetic transition [Fig.\,\ref{fig-NDb}(f)] (see also the magnetization measurements Fig.\,\ref{fig-mag}).
Yet, we found no such anomaly at 3.6\,K [Fig.\,\ref{fig-NDb}(h)], implying that at this temperature the system persists in the SDW phase in this field range.
In contrast, the measurements at 2.6\,K clearly show the existence of the satellite reflection, i.e., the indication of the spin-stripe modulation, which at 2.8(3)\,T merges with the main magnetic reflection [Fig.\,\ref{fig-NDb}(e),(g)]. 
The peak position exhibits a second anomaly at 8.0(5)\,T, which most likely indicates the transition from the VC to the SDW phase.

\subsubsection{Magnetic field perpendicular to the $b$ axis}

\begin{figure}[!]
\centering
\includegraphics[width=0.9\columnwidth]{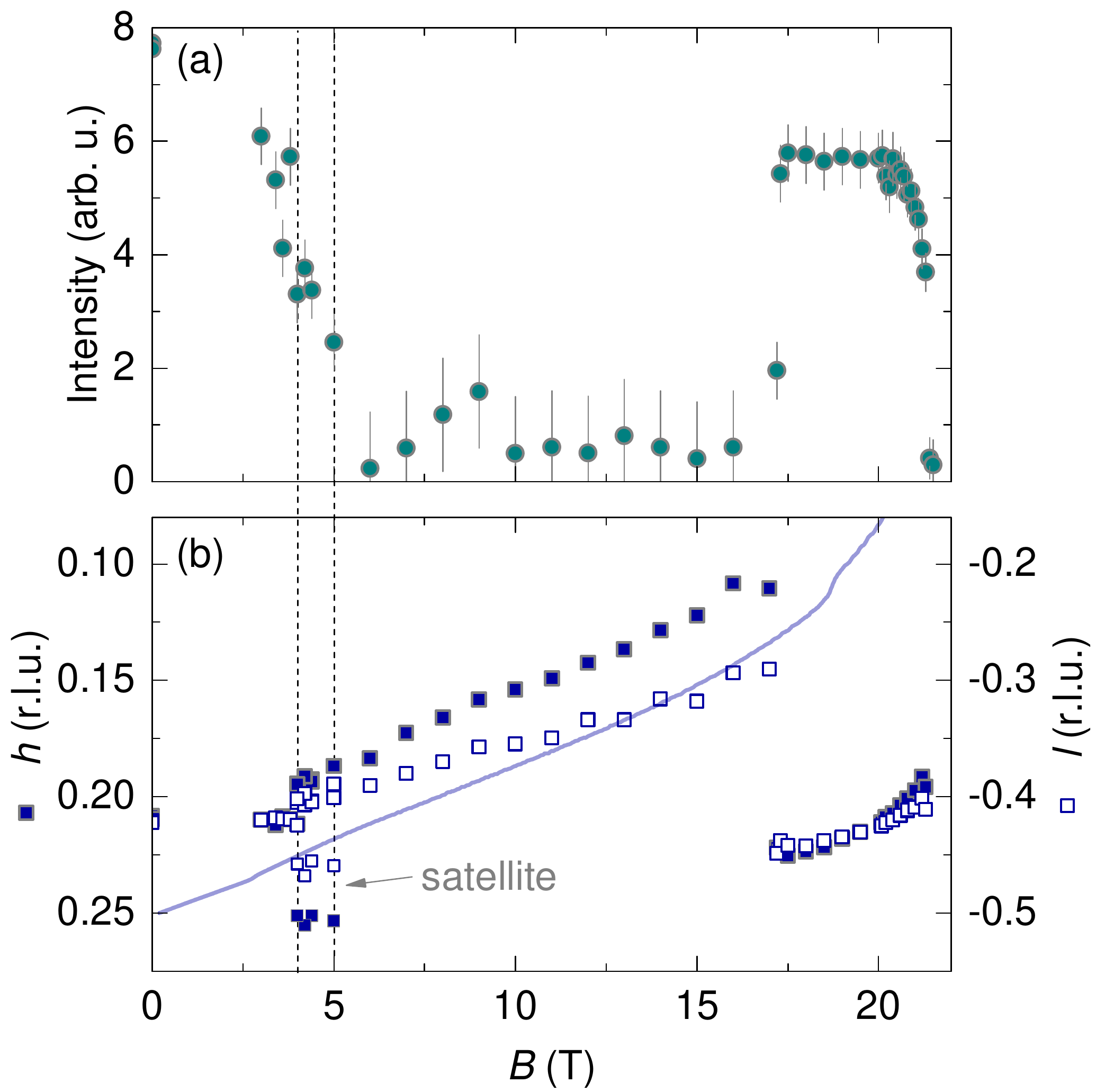}
\caption{Neutron diffraction measurements of the (0.2~0~$-$0.42) magnetic reflection at 1.7\,K for magnetic fields applied in the $ac$ plane, tilted by $\sim$30$^\circ$ from $a$ towards $c$. Panel (a) shows the integrated intensity of the magnetic reflection, while panel (b) shows the measured $l$ and $h$ components ($k$\,$\equiv$\,0) of the magnetic wave vector. The solid line indicates the theoretically predicted response for $l$ (see text for details). }
\label{fig-NDa}
\end{figure}

Neutron diffraction measurements for $B||a$ up to 6\,T have been previously reported in Ref.\,[\onlinecite{pregelj2015spin}].
We thus focus here on new experiments at even higher magnetic fields up to 25\,T.
Due to experimental limitations (see Experimental section), when probing the (0.2~0~$-$0.42) magnetic reflection the magnetic field was rotated from $a$ towards $c$ by $\sim$30$^\circ$.
However, we do not expect this rotation to have any dramatic effect, since the magnetization measurements [Fig.\,\ref{fig-mag}(a)] imply that the magnetic responses for $B||a$ and $B||c$ are similar.
In agreement with Ref.\,[\onlinecite{pregelj2015spin}] we find that increasing the magnetic field from zero does not affect the peak position, while its intensity notably decreases.
At 4.0(1)\,T, the intensity of the main magnetic reflection drops almost to zero and a weak satellite reflection emerges, which persists up to 5.0(2)\,T [Fig.\,\ref{fig-NDa}(b)], indicating the range of the spin-stripe modulation.
In parallel, the position of the main magnetic reflection shifts linearly from (0.21~0~$-$0.42) at 4\,T to (0.11~0~$-$0.29) at 17.2(1)\,T, where it jumps to (0.22~0~$-$0.44), i.e., very close to the zero-field position, and gains back the low-field intensity (Fig.\,\ref{fig-NDa}).
Further increase of the field up to 21.4\,T shifts the peak to (0.19~0~$-$0.4), which is almost exactly the zero-field position.
Eventually, at 21.4\,T, the incommensurate magnetic reflections suddenly disappear [Fig.\,\ref{fig-NDa}(b)], indicating a sharp transition into the saturated phase.
The new data thus clearly show that the HF state between 17.2 and 21.4\,T in $\beta$-TeVO$_4$ is a long-range-ordered state dominated by the incommensurate magnetic ordering.
We note that the slight deviation of the transition fields compared to the magnetization measurements for $B||a$ and $B||c$ is probably due to the fact that eigenaxes of the susceptibility tensor within the $ac$ plane are tilted by $\sim$45$^\circ$ away from the $a$ and $c$ axes.\cite{pregelj2016exchange}

\begin{figure}[!]
\centering
\includegraphics[width=\columnwidth]{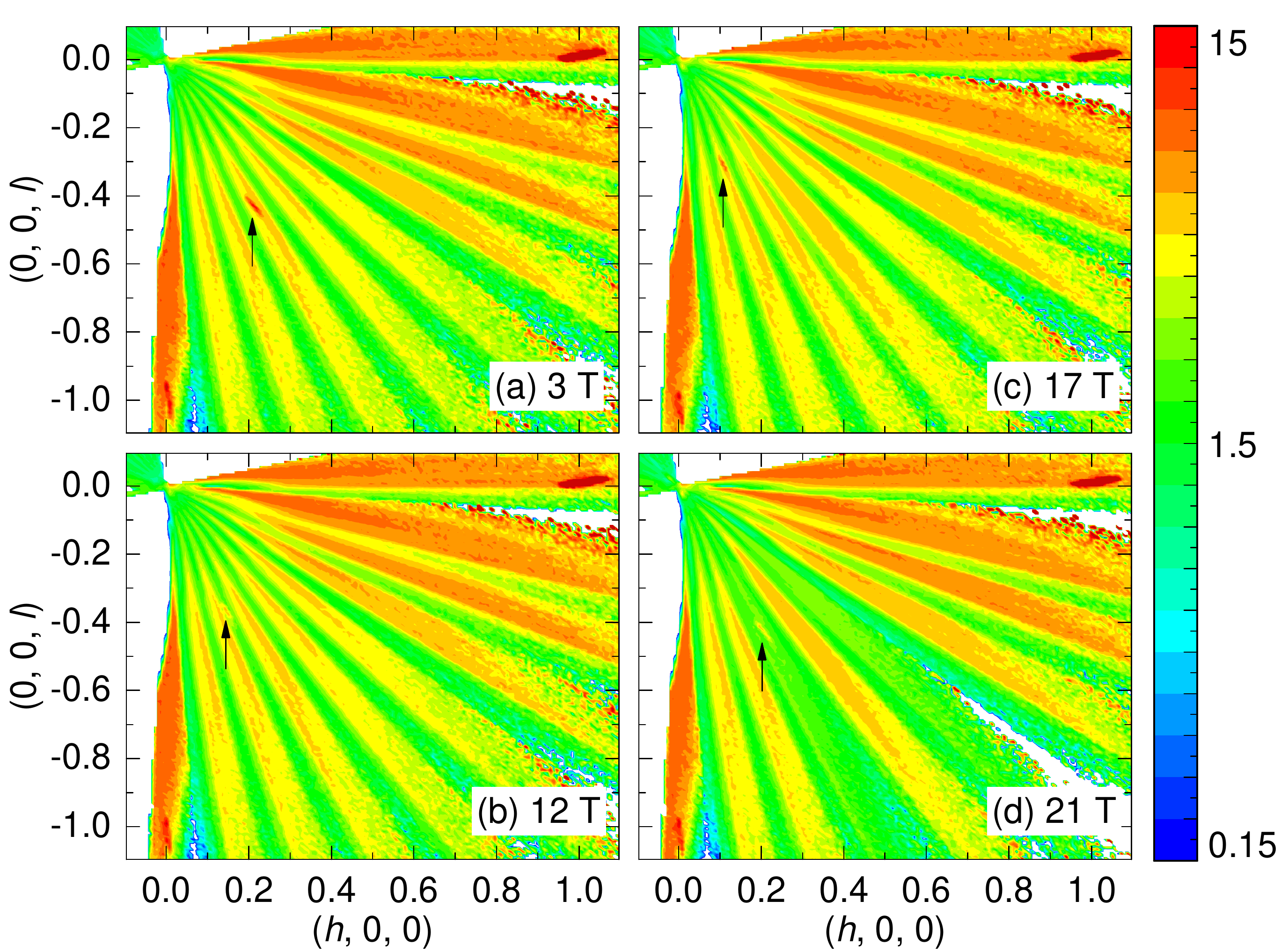}
\caption{Neutron diffraction maps of the reciprocal space at 1.7\,K for several magnetic field applied in the $ac$ plane, tilted for $\sim$30$^\circ$ from $a$ towards $c$. Panel (a) shows data collected in the VC phase at 3\,T, (b) and (c) in the SDW phase at 12\,T and 17\,T, respectively, and (d) in the HF phase at 21\,T. The arrows show the magnetic reflections. Due to experimental limitation the maps are obtained by combining data sets measured at different field orientations within the $ac$ plane.}
\label{fig-NDmaps}
\end{figure}

Finally, to check if there is any additional magnetic wave vector present in the VC (at 3\,T), SDW (at 12 and 17\,T, i.e., far above the stripe region) and HF (at 21\,T) phases, we inspected the broad region of the reciprocal space ($h$, 0, $l$), where $0\leq h \leq1$ and $-1\leq l \leq0$ (Fig.\,\ref{fig-NDmaps}).
We note that the direction of the magnetic field had to be varied within the $ac$ plane to cover the complete reciprocal-space region (see Experimental section).
Detailed inspection of the diffraction maps shows that besides the (1~0~0) and (0~0~$-$1) nuclear reflections, the only reflection that exists in this range (see arrows in Fig.\,\ref{fig-NDmaps}) is the one for which we measured complete field dependence (Fig.\,\ref{fig-NDa}).
This indicates that similarly to the VC and SDW phases the HF phase also exhibits an incommensurate magnetic order determined by a single magnetic wave vector.

\section{Discussion}

The prime goal of our study is to understand the nature of the HF phase in $\beta$-TeVO$_4$.
First, we focus on the experimentally determined magnetic phase diagrams, which we derive for magnetic fields applied along all three crystallographic axes (Fig.\,\ref{fig-PD}).
Our data are in good agreement with magnetization, heat capacity, and magnetostriction measurements presented in Ref.\,\onlinecite{weickert2016magnetic}, except for the results for $B||a$, which in Ref.\,\onlinecite{weickert2016magnetic} have probably been measured with the field applied along the $b$ axis.
We stress that the diffraction experiments do not allow for an error in the orientation of the monoclinic crystal.
Hence, we believe that responses for $B||a$ and $B||c$ are very similar, whereas for $B||b$ the low-field transition is shifted to much higher fields.
In contrast to the strong anisotropy at lower fields, the transition to the HF phase is significantly less orientation dependent (Fig.\,\ref{fig-mag}).
This suggests that the HF phase reflects the intrinsic response of the isotropic $J_1$-$J_2$ spin-1/2 chain model that predicts the quadrupolar/spin-nematic phase to develop in this magnetization region.\cite{pregelj2015spin,sudan2009emergent}
However, the presence of the magnetic reflections shows that ``conventional'' dipolar magnetic order still exists in this phase.

Another interesting result is the behavior of the magnetic wave vector.
Clearly, the latter is locked in the VC phase [Fig.\,\ref{fig-NDb}(b), Fig.\,\ref{fig-NDa}(b)], whereas it shows significant field and temperature dependencies in the SDW phase.
This is most obvious for $B$$\perp$$b$ [Fig.\,\ref{fig-NDa}(b)], where the magnetic wave vector changes almost linearly from (0.21~0~$-$0.42) at 4\,T to (0.11~0~$-$0.29) at 17.2(1)\,T.
The observed behavior is in reasonable agreement with the theoretically predicted dependence for the SDW phase in the isotropic $J_1$-$J_2$ spin-1/2 chain model, for which $l$\,=\,(1~$-M/M_{\text{sat}})/p$, where $M$ is the magnetization, $M_{\text{sat}}$ is its saturated value and $p$\,=\,2 stands for the quadrupolar spin correlations.\cite{sudan2009emergent}
On the other hand, in the HF phase the magnetic wave vector changes back to almost exactly the zero-field value, signifying a clear deviation from the isotropic $J_1$-$J_2$ spin-1/2 chain model.
In addition, we find that $h$ and $l$ components of the magnetic wave vector shift simultaneously and change by almost exactly the same amount for $B||b$ (Fig.\,\ref{fig-NDb}) as well as for $B$$\perp$$b$ [Fig.\,\ref{fig-NDa}(b)].
This suggests that despite $h$ being very close to $-l/2$, the magnetic modulation perpendicular to the chains (along $a$) is not directly determined by the modulation along the chain (along $c$).
Namely, in contrast to a single interchain interaction that would impose $h$\,=\,$-l/2$, the modulation along the $a$ axis must be affected by the competition between inter- and intra-chain interactions.
This complies with the density-functional-theory calculations,\cite{weickert2016magnetic} suggesting that there are several active ferromagnetic interchain interaction pathways along the $a$ axis, which compete with the dominant intrachain interactions.

The incommensurate magnetic order in the HF phase is most likely triggered by perturbations of the isotropic $J_1$-$J_2$ spin-1/2 chain model.
In fact, in another frustrated spin-1/2 chain candidate, linarite,\cite{willenberg2012magnetic} already small anisotropies were found to induce a fan-type spin order in applied magnetic field before the complete saturation of the magnetization was reached.\cite{cemal2018field} 
The HF state in $\beta$-TeVO$_4$ might, therefore, exhibit a similar kind of a fan order that suppresses the realization of the spin-nematic phase, predicted by the simple isotropic model.

Finally, we bring to attention studies of the spin-nematic phase in LiCuVO$_4$,\cite{svistov2011new,buttgen2014search,orlova2017nuclear} where the spin-nematic phase was suggested to exist in a very narrow range between $\sim$0.95\,$M_{\text{sat}}$ and $M_{\text{sat}}$.
In this respect, the spin-nematic phase in $\beta$-TeVO$_4$ might also develop in a few tenths of a tesla below the saturation, where neutron magnetic reflections may already vanish, while the magnetization is still not completely saturated.
The best check for this hypothesis would be precise neutron-diffraction measurements of both magnetic and nuclear reflections, the former probing the signal corresponding to the magnetic order, while the letter probing the signal of the saturated ferromagnetic phase on top of the nuclear contribution.
Alternatively, nuclear magnetic resonance measurements similar to those presented in Ref.\,\onlinecite{orlova2017nuclear} could be employed to address this intriguing question.

\section{Conclusion}

In conclusion, combining magnetization measurements in pulsed magnetic fields and neutron diffraction in static magnetic field up to 25\,T, we are able to determine in detail the magnetic phase diagram of $\beta$-TeVO$_4$.
We find a novel high-field phase that exhibits a long-range incommensurate magnetic order, which most likely suppresses spin-nematic phase due to the presence of interchain interactions and sizable magnetic anisotropies.
Nevertheless, the field dependence of the magnetic wave-vector in the SDW phase approximately follows the response predicted by the isotropic $J_1$-$J_2$ model, suggesting that the spin-nematic phase might still develop in a narrow region of a few tenths of a tesla before the full magnetic saturation is reached.

\begin{acknowledgments}

The neutron diffraction experiments were performed at the Swiss spallation neutron source SINQ, at the Paul Scherrer Institute, Villigen, Switzerland, and at Helmholtz Zentrum Berlin, Germany.
This work has been funded by the Slovenian Research Agency (project J1-9145 and program No. P1-0125) and the Swiss National Science Foundation (project SCOPES IZ73Z0\_152734/1).

\end{acknowledgments}


\begin{thebibliography}{32}%
\makeatletter
\providecommand \@ifxundefined [1]{%
 \@ifx{#1\undefined}
}%
\providecommand \@ifnum [1]{%
 \ifnum #1\expandafter \@firstoftwo
 \else \expandafter \@secondoftwo
 \fi
}%
\providecommand \@ifx [1]{%
 \ifx #1\expandafter \@firstoftwo
 \else \expandafter \@secondoftwo
 \fi
}%
\providecommand \natexlab [1]{#1}%
\providecommand \enquote  [1]{``#1''}%
\providecommand \bibnamefont  [1]{#1}%
\providecommand \bibfnamefont [1]{#1}%
\providecommand \citenamefont [1]{#1}%
\providecommand \href@noop [0]{\@secondoftwo}%
\providecommand \href [0]{\begingroup \@sanitize@url \@href}%
\providecommand \@href[1]{\@@startlink{#1}\@@href}%
\providecommand \@@href[1]{\endgroup#1\@@endlink}%
\providecommand \@sanitize@url [0]{\catcode `\\12\catcode `\$12\catcode
  `\&12\catcode `\#12\catcode `\^12\catcode `\_12\catcode `\%12\relax}%
\providecommand \@@startlink[1]{}%
\providecommand \@@endlink[0]{}%
\providecommand \url  [0]{\begingroup\@sanitize@url \@url }%
\providecommand \@url [1]{\endgroup\@href {#1}{\urlprefix }}%
\providecommand \urlprefix  [0]{URL }%
\providecommand \Eprint [0]{\href }%
\providecommand \doibase [0]{http://dx.doi.org/}%
\providecommand \selectlanguage [0]{\@gobble}%
\providecommand \bibinfo  [0]{\@secondoftwo}%
\providecommand \bibfield  [0]{\@secondoftwo}%
\providecommand \translation [1]{[#1]}%
\providecommand \BibitemOpen [0]{}%
\providecommand \bibitemStop [0]{}%
\providecommand \bibitemNoStop [0]{.\EOS\space}%
\providecommand \EOS [0]{\spacefactor3000\relax}%
\providecommand \BibitemShut  [1]{\csname bibitem#1\endcsname}%
\let\auto@bib@innerbib\@empty
\bibitem [{\citenamefont {Andreev}\ and\ \citenamefont
  {Grishchuk}(1984)}]{andreev1984spin}%
  \BibitemOpen
  \bibfield  {author} {\bibinfo {author} {\bibfnamefont {A.~F.}\ \bibnamefont
  {Andreev}}\ and\ \bibinfo {author} {\bibfnamefont {I.~A.}\ \bibnamefont
  {Grishchuk}},\ }\bibfield  {title} {\enquote {\bibinfo {title} {Spin
  nematics},}\ }\href {http://jetp.ac.ru/cgi-bin/dn/e_060_02_0267.pdf}
  {\bibfield  {journal} {\bibinfo  {journal} {Sov. Phys. JETP}\ }\textbf
  {\bibinfo {volume} {60}},\ \bibinfo {pages} {267} (\bibinfo {year}
  {1984})}\BibitemShut {NoStop}%
\bibitem [{\citenamefont {Blume}\ and\ \citenamefont
  {Hsieh}(1969)}]{blume1969biquadratic}%
  \BibitemOpen
  \bibfield  {author} {\bibinfo {author} {\bibfnamefont {M.}~\bibnamefont
  {Blume}}\ and\ \bibinfo {author} {\bibfnamefont {Y.~Y.}\ \bibnamefont
  {Hsieh}},\ }\bibfield  {title} {\enquote {\bibinfo {title} {Biquadratic
  exchange and quadrupolar ordering},}\ }\href {\doibase 10.1063/1.1657616}
  {\bibfield  {journal} {\bibinfo  {journal} {J. Appl. Phys.}\ }\textbf
  {\bibinfo {volume} {40}},\ \bibinfo {pages} {1249--1249} (\bibinfo {year}
  {1969})}\BibitemShut {NoStop}%
\bibitem [{\citenamefont {Lacroix}\ \emph {et~al.}(2011)\citenamefont
  {Lacroix}, \citenamefont {Mendels},\ and\ \citenamefont
  {Mila}}]{lacroix2011introduction}%
  \BibitemOpen
  \bibfield  {author} {\bibinfo {author} {\bibfnamefont {C.}~\bibnamefont
  {Lacroix}}, \bibinfo {author} {\bibfnamefont {P.}~\bibnamefont {Mendels}}, \
  and\ \bibinfo {author} {\bibfnamefont {F.}~\bibnamefont {Mila}},\ }\href
  {https://books.google.si/books?id=utSV09ZuhOkC} {\emph {\bibinfo {title}
  {Introduction to Frustrated Magnetism: Materials, Experiments, Theory}}},\
  Springer Series in Solid-State Sciences\ (\bibinfo  {publisher} {Springer
  Berlin Heidelberg},\ \bibinfo {year} {2011})\BibitemShut {NoStop}%
\bibitem [{\citenamefont {Hikihara}\ \emph {et~al.}(2008)\citenamefont
  {Hikihara}, \citenamefont {Kecke}, \citenamefont {Momoi},\ and\ \citenamefont
  {Furusaki}}]{hikihara2008vector}%
  \BibitemOpen
  \bibfield  {author} {\bibinfo {author} {\bibfnamefont {T.}~\bibnamefont
  {Hikihara}}, \bibinfo {author} {\bibfnamefont {L.}~\bibnamefont {Kecke}},
  \bibinfo {author} {\bibfnamefont {T.}~\bibnamefont {Momoi}}, \ and\ \bibinfo
  {author} {\bibfnamefont {A.}~\bibnamefont {Furusaki}},\ }\bibfield  {title}
  {\enquote {\bibinfo {title} {Vector chiral and multipolar orders in the
  spin-1/2 frustrated ferromagnetic chain in magnetic field},}\ }\href
  {\doibase 10.1103/PhysRevB.78.144404} {\bibfield  {journal} {\bibinfo
  {journal} {Phys. Rev. B}\ }\textbf {\bibinfo {volume} {78}},\ \bibinfo
  {pages} {144404} (\bibinfo {year} {2008})}\BibitemShut {NoStop}%
\bibitem [{\citenamefont {Sudan}\ \emph {et~al.}(2009)\citenamefont {Sudan},
  \citenamefont {L{\"u}scher},\ and\ \citenamefont
  {L{\"a}uchli}}]{sudan2009emergent}%
  \BibitemOpen
  \bibfield  {author} {\bibinfo {author} {\bibfnamefont {J.}~\bibnamefont
  {Sudan}}, \bibinfo {author} {\bibfnamefont {A.}~\bibnamefont {L{\"u}scher}},
  \ and\ \bibinfo {author} {\bibfnamefont {A.~M.}\ \bibnamefont
  {L{\"a}uchli}},\ }\bibfield  {title} {\enquote {\bibinfo {title} {{Emergent
  multipolar spin correlations in a fluctuating spiral: The frustrated
  ferromagnetic spin-1/2 Heisenberg chain in a magnetic field}},}\ }\href
  {\doibase 10.1103/PhysRevB.80.140402} {\bibfield  {journal} {\bibinfo
  {journal} {Phys. Rev. B}\ }\textbf {\bibinfo {volume} {80}},\ \bibinfo
  {pages} {140402(R)} (\bibinfo {year} {2009})}\BibitemShut {NoStop}%
\bibitem [{\citenamefont {Chubukov}(1991)}]{chubukov1991chiral}%
  \BibitemOpen
  \bibfield  {author} {\bibinfo {author} {\bibfnamefont {A.~V.}\ \bibnamefont
  {Chubukov}},\ }\bibfield  {title} {\enquote {\bibinfo {title} {Chiral,
  nematic, and dimer states in quantum spin chains},}\ }\href {\doibase
  10.1103/PhysRevB.44.4693} {\bibfield  {journal} {\bibinfo  {journal} {Phys.
  Rev. B}\ }\textbf {\bibinfo {volume} {44}},\ \bibinfo {pages} {4693}
  (\bibinfo {year} {1991})}\BibitemShut {NoStop}%
\bibitem [{\citenamefont {Shannon}\ \emph {et~al.}(2006)\citenamefont
  {Shannon}, \citenamefont {Momoi},\ and\ \citenamefont
  {Sindzingre}}]{shannon2006nematic}%
  \BibitemOpen
  \bibfield  {author} {\bibinfo {author} {\bibfnamefont {N.}~\bibnamefont
  {Shannon}}, \bibinfo {author} {\bibfnamefont {T.}~\bibnamefont {Momoi}}, \
  and\ \bibinfo {author} {\bibfnamefont {P.}~\bibnamefont {Sindzingre}},\
  }\bibfield  {title} {\enquote {\bibinfo {title} {Nematic order in square
  lattice frustrated ferromagnets},}\ }\href {\doibase
  10.1103/PhysRevLett.96.027213} {\bibfield  {journal} {\bibinfo  {journal}
  {Phys. Rev. Lett.}\ }\textbf {\bibinfo {volume} {96}},\ \bibinfo {pages}
  {027213} (\bibinfo {year} {2006})}\BibitemShut {NoStop}%
\bibitem [{\citenamefont {Zhitomirsky}\ and\ \citenamefont
  {Tsunetsugu}(2010)}]{zhitomirsky2010magnon}%
  \BibitemOpen
  \bibfield  {author} {\bibinfo {author} {\bibfnamefont {M.~E.}\ \bibnamefont
  {Zhitomirsky}}\ and\ \bibinfo {author} {\bibfnamefont {H.}~\bibnamefont
  {Tsunetsugu}},\ }\bibfield  {title} {\enquote {\bibinfo {title} {Magnon
  pairing in quantum spin nematic},}\ }\href {\doibase
  10.1209/0295-5075/92/37001} {\bibfield  {journal} {\bibinfo  {journal} {EPL}\
  }\textbf {\bibinfo {volume} {92}},\ \bibinfo {pages} {37001} (\bibinfo {year}
  {2010})}\BibitemShut {NoStop}%
\bibitem [{\citenamefont {Sato}\ \emph {et~al.}(2009)\citenamefont {Sato},
  \citenamefont {Momoi},\ and\ \citenamefont {Furusaki}}]{sato2009nmr}%
  \BibitemOpen
  \bibfield  {author} {\bibinfo {author} {\bibfnamefont {M.}~\bibnamefont
  {Sato}}, \bibinfo {author} {\bibfnamefont {T.}~\bibnamefont {Momoi}}, \ and\
  \bibinfo {author} {\bibfnamefont {A.}~\bibnamefont {Furusaki}},\ }\bibfield
  {title} {\enquote {\bibinfo {title} {{NMR relaxation rate and dynamical
  structure factors in nematic and multipolar liquids of frustrated spin chains
  under magnetic fields}},}\ }\href {\doibase 10.1103/PhysRevB.79.060406}
  {\bibfield  {journal} {\bibinfo  {journal} {Phys. Rev. B}\ }\textbf {\bibinfo
  {volume} {79}},\ \bibinfo {pages} {060406(R)} (\bibinfo {year}
  {2009})}\BibitemShut {NoStop}%
\bibitem [{\citenamefont {Starykh}\ and\ \citenamefont
  {Balents}(2014)}]{starykh2014excitations}%
  \BibitemOpen
  \bibfield  {author} {\bibinfo {author} {\bibfnamefont {O.~A.}\ \bibnamefont
  {Starykh}}\ and\ \bibinfo {author} {\bibfnamefont {L.}~\bibnamefont
  {Balents}},\ }\bibfield  {title} {\enquote {\bibinfo {title} {Excitations and
  quasi-one-dimensionality in field-induced nematic and spin density wave
  states},}\ }\href {\doibase 10.1103/PhysRevB.89.104407} {\bibfield  {journal}
  {\bibinfo  {journal} {Phys. Rev. B}\ }\textbf {\bibinfo {volume} {89}},\
  \bibinfo {pages} {104407} (\bibinfo {year} {2014})}\BibitemShut {NoStop}%
\bibitem [{\citenamefont {Orlova}\ \emph {et~al.}(2017)\citenamefont {Orlova},
  \citenamefont {Green}, \citenamefont {Law}, \citenamefont {Gorbunov},
  \citenamefont {Chanda}, \citenamefont {Kr{\"a}mer}, \citenamefont
  {Horvati{\'c}}, \citenamefont {Kremer}, \citenamefont {Wosnitza},\ and\
  \citenamefont {Rikken}}]{orlova2017nuclear}%
  \BibitemOpen
  \bibfield  {author} {\bibinfo {author} {\bibfnamefont {A.}~\bibnamefont
  {Orlova}}, \bibinfo {author} {\bibfnamefont {E.~L.}\ \bibnamefont {Green}},
  \bibinfo {author} {\bibfnamefont {J.~M.}\ \bibnamefont {Law}}, \bibinfo
  {author} {\bibfnamefont {D.~I.}\ \bibnamefont {Gorbunov}}, \bibinfo {author}
  {\bibfnamefont {G.}~\bibnamefont {Chanda}}, \bibinfo {author} {\bibfnamefont
  {S.}~\bibnamefont {Kr{\"a}mer}}, \bibinfo {author} {\bibfnamefont
  {M.}~\bibnamefont {Horvati{\'c}}}, \bibinfo {author} {\bibfnamefont {R.~K.}\
  \bibnamefont {Kremer}}, \bibinfo {author} {\bibfnamefont {J.}~\bibnamefont
  {Wosnitza}}, \ and\ \bibinfo {author} {\bibfnamefont {G.~L. J.~A.}\
  \bibnamefont {Rikken}},\ }\bibfield  {title} {\enquote {\bibinfo {title}
  {{Nuclear magnetic resonance signature of the spin-nematic phase in
  LiCuVO$_4$ at high magnetic fields}},}\ }\href {\doibase
  10.1103/PhysRevLett.118.247201} {\bibfield  {journal} {\bibinfo  {journal}
  {Phys. Rev. Lett.}\ }\textbf {\bibinfo {volume} {118}},\ \bibinfo {pages}
  {247201} (\bibinfo {year} {2017})}\BibitemShut {NoStop}%
\bibitem [{\citenamefont {Sato}\ \emph {et~al.}(2011)\citenamefont {Sato},
  \citenamefont {Hikihara},\ and\ \citenamefont {Momoi}}]{sato2011field}%
  \BibitemOpen
  \bibfield  {author} {\bibinfo {author} {\bibfnamefont {M.}~\bibnamefont
  {Sato}}, \bibinfo {author} {\bibfnamefont {T.}~\bibnamefont {Hikihara}}, \
  and\ \bibinfo {author} {\bibfnamefont {T.}~\bibnamefont {Momoi}},\ }\bibfield
   {title} {\enquote {\bibinfo {title} {{Field and temperature dependence of
  NMR relaxation rate in the magnetic quadrupolar liquid phase of spin-1/2
  frustrated ferromagnetic chains}},}\ }\href {\doibase
  10.1103/PhysRevB.83.064405} {\bibfield  {journal} {\bibinfo  {journal} {Phys.
  Rev. B}\ }\textbf {\bibinfo {volume} {83}},\ \bibinfo {pages} {064405}
  (\bibinfo {year} {2011})}\BibitemShut {NoStop}%
\bibitem [{\citenamefont {Onishi}(2015)}]{onishi2015magnetic}%
  \BibitemOpen
  \bibfield  {author} {\bibinfo {author} {\bibfnamefont {H.}~\bibnamefont
  {Onishi}},\ }\bibfield  {title} {\enquote {\bibinfo {title} {Magnetic
  excitations of spin nematic state in frustrated ferromagnetic chain},}\
  }\href {\doibase 10.7566/JPSJ.84.083702} {\bibfield  {journal} {\bibinfo
  {journal} {J. Phys. Soc. Jap.}\ }\textbf {\bibinfo {volume} {84}},\ \bibinfo
  {pages} {083702} (\bibinfo {year} {2015})}\BibitemShut {NoStop}%
\bibitem [{\citenamefont {Furuya}(2017)}]{furuya2017angular}%
  \BibitemOpen
  \bibfield  {author} {\bibinfo {author} {\bibfnamefont {S.~C.}\ \bibnamefont
  {Furuya}},\ }\bibfield  {title} {\enquote {\bibinfo {title} {Angular
  dependence of electron spin resonance for detecting the quadrupolar liquid
  state of frustrated spin chains},}\ }\href {\doibase
  10.1103/PhysRevB.95.014416} {\bibfield  {journal} {\bibinfo  {journal} {Phys.
  Rev. B}\ }\textbf {\bibinfo {volume} {95}},\ \bibinfo {pages} {014416}
  (\bibinfo {year} {2017})}\BibitemShut {NoStop}%
\bibitem [{\citenamefont {Svistov}\ \emph {et~al.}(2011)\citenamefont
  {Svistov}, \citenamefont {Fujita}, \citenamefont {Yamaguchi}, \citenamefont
  {Kimura}, \citenamefont {Omura}, \citenamefont {Prokofiev}, \citenamefont
  {Smirnov}, \citenamefont {Honda},\ and\ \citenamefont
  {Hagiwara}}]{svistov2011new}%
  \BibitemOpen
  \bibfield  {author} {\bibinfo {author} {\bibfnamefont {L.~E.}\ \bibnamefont
  {Svistov}}, \bibinfo {author} {\bibfnamefont {T.}~\bibnamefont {Fujita}},
  \bibinfo {author} {\bibfnamefont {H.}~\bibnamefont {Yamaguchi}}, \bibinfo
  {author} {\bibfnamefont {S.}~\bibnamefont {Kimura}}, \bibinfo {author}
  {\bibfnamefont {K.}~\bibnamefont {Omura}}, \bibinfo {author} {\bibfnamefont
  {A.}~\bibnamefont {Prokofiev}}, \bibinfo {author} {\bibfnamefont {A.~I.}\
  \bibnamefont {Smirnov}}, \bibinfo {author} {\bibfnamefont {Z.}~\bibnamefont
  {Honda}}, \ and\ \bibinfo {author} {\bibfnamefont {M.}~\bibnamefont
  {Hagiwara}},\ }\bibfield  {title} {\enquote {\bibinfo {title} {{New high
  magnetic field phase of the frustrated $S$=1/2 chain compound LiCuVO$_4$}},}\
  }\href {\doibase 10.1134/S0021364011010073} {\bibfield  {journal} {\bibinfo
  {journal} {JETP letters}\ }\textbf {\bibinfo {volume} {93}},\ \bibinfo
  {pages} {21--25} (\bibinfo {year} {2011})}\BibitemShut {NoStop}%
\bibitem [{\citenamefont {Mourigal}\ \emph {et~al.}(2012)\citenamefont
  {Mourigal}, \citenamefont {Enderle}, \citenamefont {F{\aa}k}, \citenamefont
  {Kremer}, \citenamefont {Law}, \citenamefont {Schneidewind}, \citenamefont
  {Hiess},\ and\ \citenamefont {Prokofiev}}]{mourigal2012evidence}%
  \BibitemOpen
  \bibfield  {author} {\bibinfo {author} {\bibfnamefont {M.}~\bibnamefont
  {Mourigal}}, \bibinfo {author} {\bibfnamefont {M.}~\bibnamefont {Enderle}},
  \bibinfo {author} {\bibfnamefont {B.}~\bibnamefont {F{\aa}k}}, \bibinfo
  {author} {\bibfnamefont {R.~K.}\ \bibnamefont {Kremer}}, \bibinfo {author}
  {\bibfnamefont {J.~M.}\ \bibnamefont {Law}}, \bibinfo {author} {\bibfnamefont
  {A.}~\bibnamefont {Schneidewind}}, \bibinfo {author} {\bibfnamefont
  {A.}~\bibnamefont {Hiess}}, \ and\ \bibinfo {author} {\bibfnamefont
  {A.}~\bibnamefont {Prokofiev}},\ }\bibfield  {title} {\enquote {\bibinfo
  {title} {{Evidence of a bond-nematic phase in LiCuVO$_4$}},}\ }\href
  {\doibase 10.1103/PhysRevLett.109.027203} {\bibfield  {journal} {\bibinfo
  {journal} {Phys. Rev. Lett.}\ }\textbf {\bibinfo {volume} {109}},\ \bibinfo
  {pages} {027203} (\bibinfo {year} {2012})}\BibitemShut {NoStop}%
\bibitem [{\citenamefont {B{\"u}ttgen}\ \emph {et~al.}(2014)\citenamefont
  {B{\"u}ttgen}, \citenamefont {Nawa}, \citenamefont {Fujita}, \citenamefont
  {Hagiwara}, \citenamefont {Kuhns}, \citenamefont {Prokofiev}, \citenamefont
  {Reyes}, \citenamefont {Svistov}, \citenamefont {Yoshimura},\ and\
  \citenamefont {Takigawa}}]{buttgen2014search}%
  \BibitemOpen
  \bibfield  {author} {\bibinfo {author} {\bibfnamefont {N.}~\bibnamefont
  {B{\"u}ttgen}}, \bibinfo {author} {\bibfnamefont {K.}~\bibnamefont {Nawa}},
  \bibinfo {author} {\bibfnamefont {T.}~\bibnamefont {Fujita}}, \bibinfo
  {author} {\bibfnamefont {M.}~\bibnamefont {Hagiwara}}, \bibinfo {author}
  {\bibfnamefont {P.}~\bibnamefont {Kuhns}}, \bibinfo {author} {\bibfnamefont
  {A.}~\bibnamefont {Prokofiev}}, \bibinfo {author} {\bibfnamefont {A.~P.}\
  \bibnamefont {Reyes}}, \bibinfo {author} {\bibfnamefont {L.~E.}\ \bibnamefont
  {Svistov}}, \bibinfo {author} {\bibfnamefont {K.}~\bibnamefont {Yoshimura}},
  \ and\ \bibinfo {author} {\bibfnamefont {M.}~\bibnamefont {Takigawa}},\
  }\bibfield  {title} {\enquote {\bibinfo {title} {{Search for a spin-nematic
  phase in the quasi-one-dimensional frustrated magnet LiCuVO$_4$}},}\ }\href
  {\doibase 10.1103/PhysRevB.90.134401} {\bibfield  {journal} {\bibinfo
  {journal} {Phys. Rev. B}\ }\textbf {\bibinfo {volume} {90}},\ \bibinfo
  {pages} {134401} (\bibinfo {year} {2014})}\BibitemShut {NoStop}%
\bibitem [{\citenamefont {Meunier}\ \emph {et~al.}(1973)\citenamefont
  {Meunier}, \citenamefont {Darriet},\ and\ \citenamefont
  {Galy}}]{meunier1973oxyde}%
  \BibitemOpen
  \bibfield  {author} {\bibinfo {author} {\bibfnamefont {G.}~\bibnamefont
  {Meunier}}, \bibinfo {author} {\bibfnamefont {J.}~\bibnamefont {Darriet}}, \
  and\ \bibinfo {author} {\bibfnamefont {J.}~\bibnamefont {Galy}},\ }\bibfield
  {title} {\enquote {\bibinfo {title} {{L'oxyde double TeVO$_4$ II. Structure
  cristalline de TeVO$_4$-$\beta$-relations structurales}},}\ }\href {\doibase
  10.1016/0022-4596(73)90205-3} {\bibfield  {journal} {\bibinfo  {journal} {J.
  Sol. Stat. Chem.}\ }\textbf {\bibinfo {volume} {6}},\ \bibinfo {pages}
  {67--73} (\bibinfo {year} {1973})}\BibitemShut {NoStop}%
\bibitem [{\citenamefont {Savina}\ \emph {et~al.}(2011)\citenamefont {Savina},
  \citenamefont {Bludov}, \citenamefont {Pashchenko}, \citenamefont
  {Gnatchenko}, \citenamefont {Lemmens},\ and\ \citenamefont
  {Berger}}]{savina2011magnetic}%
  \BibitemOpen
  \bibfield  {author} {\bibinfo {author} {\bibfnamefont {Yu}~\bibnamefont
  {Savina}}, \bibinfo {author} {\bibfnamefont {O.}~\bibnamefont {Bludov}},
  \bibinfo {author} {\bibfnamefont {V.}~\bibnamefont {Pashchenko}}, \bibinfo
  {author} {\bibfnamefont {S.L.}\ \bibnamefont {Gnatchenko}}, \bibinfo {author}
  {\bibfnamefont {P.}~\bibnamefont {Lemmens}}, \ and\ \bibinfo {author}
  {\bibfnamefont {H.}~\bibnamefont {Berger}},\ }\bibfield  {title} {\enquote
  {\bibinfo {title} {{Magnetic properties of the antiferromagnetic spin-1/2
  chain system $\beta$-TeVO$_4$}},}\ }\href {\doibase
  10.1103/PhysRevB.84.104447} {\bibfield  {journal} {\bibinfo  {journal} {Phys.
  Rev. B}\ }\textbf {\bibinfo {volume} {84}},\ \bibinfo {pages} {104447}
  (\bibinfo {year} {2011})}\BibitemShut {NoStop}%
\bibitem [{\citenamefont {Pregelj}\ \emph {et~al.}(2015)\citenamefont
  {Pregelj}, \citenamefont {Zorko}, \citenamefont {Zaharko}, \citenamefont
  {Nojiri}, \citenamefont {Berger}, \citenamefont {Chapon},\ and\ \citenamefont
  {Ar{\v{c}}on}}]{pregelj2015spin}%
  \BibitemOpen
  \bibfield  {author} {\bibinfo {author} {\bibfnamefont {M.}~\bibnamefont
  {Pregelj}}, \bibinfo {author} {\bibfnamefont {A.}~\bibnamefont {Zorko}},
  \bibinfo {author} {\bibfnamefont {O.}~\bibnamefont {Zaharko}}, \bibinfo
  {author} {\bibfnamefont {H.}~\bibnamefont {Nojiri}}, \bibinfo {author}
  {\bibfnamefont {H.}~\bibnamefont {Berger}}, \bibinfo {author} {\bibfnamefont
  {L.C.}\ \bibnamefont {Chapon}}, \ and\ \bibinfo {author} {\bibfnamefont
  {D.}~\bibnamefont {Ar{\v{c}}on}},\ }\bibfield  {title} {\enquote {\bibinfo
  {title} {{Spin-stripe phase in a frustrated zigzag spin-1/2 chain}},}\ }\href
  {\doibase 10.1038/ncomms8255} {\bibfield  {journal} {\bibinfo  {journal}
  {Nat. Commun.}\ }\textbf {\bibinfo {volume} {6}},\ \bibinfo {pages} {7255}
  (\bibinfo {year} {2015})}\BibitemShut {NoStop}%
\bibitem [{\citenamefont {Savina}\ \emph {et~al.}(2015)\citenamefont {Savina},
  \citenamefont {Bludov}, \citenamefont {Pashchenko}, \citenamefont
  {Gnatchenko}, \citenamefont {Savin}, \citenamefont {Sch{\"a}fer},
  \citenamefont {Lemmens},\ and\ \citenamefont {Berger}}]{savina2015study}%
  \BibitemOpen
  \bibfield  {author} {\bibinfo {author} {\bibfnamefont {Yu~O.}\ \bibnamefont
  {Savina}}, \bibinfo {author} {\bibfnamefont {A.~N.}\ \bibnamefont {Bludov}},
  \bibinfo {author} {\bibfnamefont {V.~A.}\ \bibnamefont {Pashchenko}},
  \bibinfo {author} {\bibfnamefont {S.~L.}\ \bibnamefont {Gnatchenko}},
  \bibinfo {author} {\bibfnamefont {Yu~V.}\ \bibnamefont {Savin}}, \bibinfo
  {author} {\bibfnamefont {S.}~\bibnamefont {Sch{\"a}fer}}, \bibinfo {author}
  {\bibfnamefont {P.}~\bibnamefont {Lemmens}}, \ and\ \bibinfo {author}
  {\bibfnamefont {H.}~\bibnamefont {Berger}},\ }\bibfield  {title} {\enquote
  {\bibinfo {title} {{A study of the magnetic properties of a
  quasi-one-dimensional magnet $\beta$-TeVO$_4$ in the frame of the
  $J_1$--$J_2$ model}},}\ }\href {\doibase 10.1063/1.4928922} {\bibfield
  {journal} {\bibinfo  {journal} {Low Temp. Phys.}\ }\textbf {\bibinfo {volume}
  {41}},\ \bibinfo {pages} {659--661} (\bibinfo {year} {2015})}\BibitemShut
  {NoStop}%
\bibitem [{\citenamefont {Weickert}\ \emph {et~al.}(2016)\citenamefont
  {Weickert}, \citenamefont {Harrison}, \citenamefont {Scott}, \citenamefont
  {Jaime}, \citenamefont {Leitm{\"a}e}, \citenamefont {Heinmaa}, \citenamefont
  {Stern}, \citenamefont {Janson}, \citenamefont {Berger}, \citenamefont
  {Rosner} \emph {et~al.}}]{weickert2016magnetic}%
  \BibitemOpen
  \bibfield  {author} {\bibinfo {author} {\bibfnamefont {F.}~\bibnamefont
  {Weickert}}, \bibinfo {author} {\bibfnamefont {N.}~\bibnamefont {Harrison}},
  \bibinfo {author} {\bibfnamefont {B.~L.}\ \bibnamefont {Scott}}, \bibinfo
  {author} {\bibfnamefont {M.}~\bibnamefont {Jaime}}, \bibinfo {author}
  {\bibfnamefont {A.}~\bibnamefont {Leitm{\"a}e}}, \bibinfo {author}
  {\bibfnamefont {I.}~\bibnamefont {Heinmaa}}, \bibinfo {author} {\bibfnamefont
  {R.}~\bibnamefont {Stern}}, \bibinfo {author} {\bibfnamefont
  {O.}~\bibnamefont {Janson}}, \bibinfo {author} {\bibfnamefont
  {H.}~\bibnamefont {Berger}}, \bibinfo {author} {\bibfnamefont
  {H.}~\bibnamefont {Rosner}},  \emph {et~al.},\ }\bibfield  {title} {\enquote
  {\bibinfo {title} {{Magnetic anisotropy in the frustrated spin-chain compound
  $\beta$-TeVO$_4$}},}\ }\href {\doibase 10.1103/PhysRevB.94.064403} {\bibfield
   {journal} {\bibinfo  {journal} {Phys. Rev. B}\ }\textbf {\bibinfo {volume}
  {94}},\ \bibinfo {pages} {064403} (\bibinfo {year} {2016})}\BibitemShut
  {NoStop}%
\bibitem [{\citenamefont {Pregelj}\ \emph {et~al.}(2018)\citenamefont
  {Pregelj}, \citenamefont {Zaharko}, \citenamefont {Stuhr}, \citenamefont
  {Zorko}, \citenamefont {Berger}, \citenamefont {Prokofiev},\ and\
  \citenamefont {Ar{\v{c}}on}}]{pregelj2018coexisting}%
  \BibitemOpen
  \bibfield  {author} {\bibinfo {author} {\bibfnamefont {M.}~\bibnamefont
  {Pregelj}}, \bibinfo {author} {\bibfnamefont {O.}~\bibnamefont {Zaharko}},
  \bibinfo {author} {\bibfnamefont {U.}~\bibnamefont {Stuhr}}, \bibinfo
  {author} {\bibfnamefont {A.}~\bibnamefont {Zorko}}, \bibinfo {author}
  {\bibfnamefont {H.}~\bibnamefont {Berger}}, \bibinfo {author} {\bibfnamefont
  {A.}~\bibnamefont {Prokofiev}}, \ and\ \bibinfo {author} {\bibfnamefont
  {D.}~\bibnamefont {Ar{\v{c}}on}},\ }\bibfield  {title} {\enquote {\bibinfo
  {title} {{Coexisting spinons and magnons in frustrated zigzag spin-1/2 chain
  compound $\beta$-TeVO$_4$}},}\ }\href {\doibase 10.1103/PhysRevB.98.094405}
  {\bibfield  {journal} {\bibinfo  {journal} {Phys. Rev. B}\ }\textbf {\bibinfo
  {volume} {98}},\ \bibinfo {pages} {094405} (\bibinfo {year}
  {2018})}\BibitemShut {NoStop}%
\bibitem [{\citenamefont {Prokhnenko}\ \emph {et~al.}(2015)\citenamefont
  {Prokhnenko}, \citenamefont {Stein}, \citenamefont {Bleif}, \citenamefont
  {Fromme}, \citenamefont {Bartkowiak},\ and\ \citenamefont
  {Wilpert}}]{prokhnenko2015time}%
  \BibitemOpen
  \bibfield  {author} {\bibinfo {author} {\bibfnamefont {O.}~\bibnamefont
  {Prokhnenko}}, \bibinfo {author} {\bibfnamefont {W.-D.}\ \bibnamefont
  {Stein}}, \bibinfo {author} {\bibfnamefont {H.-J.}\ \bibnamefont {Bleif}},
  \bibinfo {author} {\bibfnamefont {M.}~\bibnamefont {Fromme}}, \bibinfo
  {author} {\bibfnamefont {M.}~\bibnamefont {Bartkowiak}}, \ and\ \bibinfo
  {author} {\bibfnamefont {T.}~\bibnamefont {Wilpert}},\ }\bibfield  {title}
  {\enquote {\bibinfo {title} {{Time-of-flight extreme environment
  diffractometer at the Helmholtz-Zentrum Berlin}},}\ }\href {\doibase
  10.1063/1.4913656} {\bibfield  {journal} {\bibinfo  {journal} {Rev. Sci.
  Instrum.}\ }\textbf {\bibinfo {volume} {86}},\ \bibinfo {pages} {033102}
  (\bibinfo {year} {2015})}\BibitemShut {NoStop}%
\bibitem [{\citenamefont {Prokhnenko}\ \emph {et~al.}(2017)\citenamefont
  {Prokhnenko}, \citenamefont {Smeibidl}, \citenamefont {Stein}, \citenamefont
  {Bartkowiak},\ and\ \citenamefont {St{\"u}sser}}]{prokhnenko2017hfm}%
  \BibitemOpen
  \bibfield  {author} {\bibinfo {author} {\bibfnamefont {O.}~\bibnamefont
  {Prokhnenko}}, \bibinfo {author} {\bibfnamefont {P.}~\bibnamefont
  {Smeibidl}}, \bibinfo {author} {\bibfnamefont {W.-D.}\ \bibnamefont {Stein}},
  \bibinfo {author} {\bibfnamefont {M.}~\bibnamefont {Bartkowiak}}, \ and\
  \bibinfo {author} {\bibfnamefont {N.}~\bibnamefont {St{\"u}sser}},\
  }\bibfield  {title} {\enquote {\bibinfo {title} {{HFM/EXED: The High Magnetic
  Field Facility for Neutron Scattering at BER II}},}\ }\href {\doibase
  10.17815/jlsrf-3-111} {\bibfield  {journal} {\bibinfo  {journal} {Journal of
  large-scale research facilities JLSRF}\ }\textbf {\bibinfo {volume} {3}},\
  \bibinfo {pages} {115} (\bibinfo {year} {2017})}\BibitemShut {NoStop}%
\bibitem [{\citenamefont {Pregelj}\ \emph {et~al.}(2016)\citenamefont
  {Pregelj}, \citenamefont {Zaharko}, \citenamefont {Herak}, \citenamefont
  {Gomil{\v{s}}ek}, \citenamefont {Zorko}, \citenamefont {Chapon},
  \citenamefont {Bourdarot}, \citenamefont {Berger},\ and\ \citenamefont
  {Ar{\v{c}}on}}]{pregelj2016exchange}%
  \BibitemOpen
  \bibfield  {author} {\bibinfo {author} {\bibfnamefont {M.}~\bibnamefont
  {Pregelj}}, \bibinfo {author} {\bibfnamefont {O.}~\bibnamefont {Zaharko}},
  \bibinfo {author} {\bibfnamefont {M.}~\bibnamefont {Herak}}, \bibinfo
  {author} {\bibfnamefont {M.}~\bibnamefont {Gomil{\v{s}}ek}}, \bibinfo
  {author} {\bibfnamefont {A.}~\bibnamefont {Zorko}}, \bibinfo {author}
  {\bibfnamefont {L.~C.}\ \bibnamefont {Chapon}}, \bibinfo {author}
  {\bibfnamefont {F.}~\bibnamefont {Bourdarot}}, \bibinfo {author}
  {\bibfnamefont {H.}~\bibnamefont {Berger}}, \ and\ \bibinfo {author}
  {\bibfnamefont {D.}~\bibnamefont {Ar{\v{c}}on}},\ }\bibfield  {title}
  {\enquote {\bibinfo {title} {{Exchange anisotropy as mechanism for
  spin-stripe formation in frustrated spin chains}},}\ }\href {\doibase
  10.1103/PhysRevB.94.081114} {\bibfield  {journal} {\bibinfo  {journal} {Phys.
  Rev. B}\ }\textbf {\bibinfo {volume} {94}},\ \bibinfo {pages} {081114(R)}
  (\bibinfo {year} {2016})}\BibitemShut {NoStop}%
\bibitem [{\citenamefont {Pregelj}\ \emph {et~al.}(2019)\citenamefont
  {Pregelj}, \citenamefont {Zorko}, \citenamefont {Gomil{\v{s}}ek},
  \citenamefont {Klanj{\v{s}}ek}, \citenamefont {Zaharko}, \citenamefont
  {White}, \citenamefont {Luetkens}, \citenamefont {Coomer}, \citenamefont
  {Ivek}, \citenamefont {G{\'o}ngora}, \citenamefont {Berger},\ and\
  \citenamefont {Ar{\v{c}}on}}]{pregelj2019elementary}%
  \BibitemOpen
  \bibfield  {author} {\bibinfo {author} {\bibfnamefont {M.}~\bibnamefont
  {Pregelj}}, \bibinfo {author} {\bibfnamefont {A.}~\bibnamefont {Zorko}},
  \bibinfo {author} {\bibfnamefont {M.}~\bibnamefont {Gomil{\v{s}}ek}},
  \bibinfo {author} {\bibfnamefont {M.}~\bibnamefont {Klanj{\v{s}}ek}},
  \bibinfo {author} {\bibfnamefont {O.}~\bibnamefont {Zaharko}}, \bibinfo
  {author} {\bibfnamefont {J.S.}\ \bibnamefont {White}}, \bibinfo {author}
  {\bibfnamefont {H.}~\bibnamefont {Luetkens}}, \bibinfo {author}
  {\bibfnamefont {F.}~\bibnamefont {Coomer}}, \bibinfo {author} {\bibfnamefont
  {T.}~\bibnamefont {Ivek}}, \bibinfo {author} {\bibfnamefont {D.R.}\
  \bibnamefont {G{\'o}ngora}}, \bibinfo {author} {\bibfnamefont
  {H.}~\bibnamefont {Berger}}, \ and\ \bibinfo {author} {\bibfnamefont
  {D.}~\bibnamefont {Ar{\v{c}}on}},\ }\bibfield  {title} {\enquote {\bibinfo
  {title} {Elementary excitation in the spin-stripe phase in quantum chains},}\
  }\href {\doibase 10.1038/s41535-019-0160-5} {\bibfield  {journal} {\bibinfo
  {journal} {npj Quantum Mater.}\ }\textbf {\bibinfo {volume} {4}},\ \bibinfo
  {pages} {22} (\bibinfo {year} {2019})}\BibitemShut {NoStop}%
\bibitem [{\citenamefont {Taroni}\ \emph {et~al.}(2008)\citenamefont {Taroni},
  \citenamefont {Bramwell},\ and\ \citenamefont
  {Holdsworth}}]{taroni2008universal}%
  \BibitemOpen
  \bibfield  {author} {\bibinfo {author} {\bibfnamefont {A.}~\bibnamefont
  {Taroni}}, \bibinfo {author} {\bibfnamefont {S.~T.}\ \bibnamefont
  {Bramwell}}, \ and\ \bibinfo {author} {\bibfnamefont {P.~C.~W.}\ \bibnamefont
  {Holdsworth}},\ }\bibfield  {title} {\enquote {\bibinfo {title} {Universal
  window for two-dimensional critical exponents},}\ }\href {\doibase
  10.1088/0953-8984/20/27/275233} {\bibfield  {journal} {\bibinfo  {journal}
  {J. Phys. Condens. Matter}\ }\textbf {\bibinfo {volume} {20}},\ \bibinfo
  {pages} {275233} (\bibinfo {year} {2008})}\BibitemShut {NoStop}%
\bibitem [{\citenamefont {Pelissetto}\ and\ \citenamefont
  {Vicari}(2002)}]{pelissetto2002critical}%
  \BibitemOpen
  \bibfield  {author} {\bibinfo {author} {\bibfnamefont {Andrea}\ \bibnamefont
  {Pelissetto}}\ and\ \bibinfo {author} {\bibfnamefont {Ettore}\ \bibnamefont
  {Vicari}},\ }\bibfield  {title} {\enquote {\bibinfo {title} {Critical
  phenomena and renormalization-group theory},}\ }\href {\doibase
  10.1016/S0370-1573(02)00219-3} {\bibfield  {journal} {\bibinfo  {journal}
  {Phys. Rep.}\ }\textbf {\bibinfo {volume} {368}},\ \bibinfo {pages}
  {549--727} (\bibinfo {year} {2002})}\BibitemShut {NoStop}%
\bibitem [{\citenamefont {Chaikin}\ and\ \citenamefont
  {Lubensky}(2000)}]{chaikin2000principles}%
  \BibitemOpen
  \bibfield  {author} {\bibinfo {author} {\bibfnamefont {P.~M.}\ \bibnamefont
  {Chaikin}}\ and\ \bibinfo {author} {\bibfnamefont {T.~C.}\ \bibnamefont
  {Lubensky}},\ }\href@noop {} {\emph {\bibinfo {title} {Principles of
  condensed matter physics}}},\ Vol.~\bibinfo {volume} {1}\ (\bibinfo
  {publisher} {Cambridge university press Cambridge},\ \bibinfo {year}
  {2000})\BibitemShut {NoStop}%
\bibitem [{\citenamefont {Willenberg}\ \emph {et~al.}(2012)\citenamefont
  {Willenberg}, \citenamefont {Sch{\"a}pers}, \citenamefont {Rule},
  \citenamefont {S{\"u}llow}, \citenamefont {Reehuis}, \citenamefont {Ryll},
  \citenamefont {Klemke}, \citenamefont {Kiefer}, \citenamefont
  {Schottenhamel}, \citenamefont {B{\"u}chner} \emph
  {et~al.}}]{willenberg2012magnetic}%
  \BibitemOpen
  \bibfield  {author} {\bibinfo {author} {\bibfnamefont {B.}~\bibnamefont
  {Willenberg}}, \bibinfo {author} {\bibfnamefont {M.}~\bibnamefont
  {Sch{\"a}pers}}, \bibinfo {author} {\bibfnamefont {K.~C.}\ \bibnamefont
  {Rule}}, \bibinfo {author} {\bibfnamefont {S.}~\bibnamefont {S{\"u}llow}},
  \bibinfo {author} {\bibfnamefont {M.}~\bibnamefont {Reehuis}}, \bibinfo
  {author} {\bibfnamefont {H.}~\bibnamefont {Ryll}}, \bibinfo {author}
  {\bibfnamefont {B.}~\bibnamefont {Klemke}}, \bibinfo {author} {\bibfnamefont
  {K.}~\bibnamefont {Kiefer}}, \bibinfo {author} {\bibfnamefont
  {W.}~\bibnamefont {Schottenhamel}}, \bibinfo {author} {\bibfnamefont
  {B.}~\bibnamefont {B{\"u}chner}},  \emph {et~al.},\ }\bibfield  {title}
  {\enquote {\bibinfo {title} {{Magnetic frustration in a quantum spin chain:
  The case of linarite PbCuSo$_4$(OH)$_2$}},}\ }\href {\doibase
  10.1103/PhysRevLett.108.117202} {\bibfield  {journal} {\bibinfo  {journal}
  {Phys. Rev. Lett.}\ }\textbf {\bibinfo {volume} {108}},\ \bibinfo {pages}
  {117202} (\bibinfo {year} {2012})}\BibitemShut {NoStop}%
\bibitem [{\citenamefont {Cemal}\ \emph {et~al.}(2018)\citenamefont {Cemal},
  \citenamefont {Enderle}, \citenamefont {Kremer}, \citenamefont {F{\aa}k},
  \citenamefont {Ressouche}, \citenamefont {Goff}, \citenamefont {Gvozdikova},
  \citenamefont {Zhitomirsky},\ and\ \citenamefont {Ziman}}]{cemal2018field}%
  \BibitemOpen
  \bibfield  {author} {\bibinfo {author} {\bibfnamefont {E.}~\bibnamefont
  {Cemal}}, \bibinfo {author} {\bibfnamefont {M.}~\bibnamefont {Enderle}},
  \bibinfo {author} {\bibfnamefont {R.~K.}\ \bibnamefont {Kremer}}, \bibinfo
  {author} {\bibfnamefont {B.}~\bibnamefont {F{\aa}k}}, \bibinfo {author}
  {\bibfnamefont {E.}~\bibnamefont {Ressouche}}, \bibinfo {author}
  {\bibfnamefont {J.~P.}\ \bibnamefont {Goff}}, \bibinfo {author}
  {\bibfnamefont {M.~V.}\ \bibnamefont {Gvozdikova}}, \bibinfo {author}
  {\bibfnamefont {M.~E.}\ \bibnamefont {Zhitomirsky}}, \ and\ \bibinfo {author}
  {\bibfnamefont {T.}~\bibnamefont {Ziman}},\ }\bibfield  {title} {\enquote
  {\bibinfo {title} {Field-induced states and excitations in the quasicritical
  spin-1/2 chain linarite},}\ }\href {\doibase 10.1103/PhysRevLett.120.067203}
  {\bibfield  {journal} {\bibinfo  {journal} {Phys. Rev. Lett.}\ }\textbf
  {\bibinfo {volume} {120}},\ \bibinfo {pages} {067203} (\bibinfo {year}
  {2018})}\BibitemShut {NoStop}%
\end{thebibliography}
\end{document}